\documentclass[twoside]{article}      
\usepackage{amsmath,amssymb,amsfonts,amsthm} 
\usepackage{graphics}                 
\usepackage{color}                    
\usepackage{hyperref}                 
\usepackage{comment} 			
\usepackage{enumitem}
\usepackage{tikz}

\theoremstyle{plain}
\newtheorem{theorem}{Theorem}[section]
\newtheorem{lemma}[theorem]{Lemma}

\newtheorem{conjj}[theorem]{Conjecture}

\theoremstyle{definition}
\newtheorem{definition}{Definition}[section]

\theoremstyle{definition}
\newtheorem{exmp}{Example}[section]

\def\B{\mathbb{ B}}

\def\C{\mathfrak{C}}

\date{\small\it \today}
\title{Sampling and Complexity of Partition Function
\footnote{Great thanks for whole heart support of my wife. Thanks for Internet and research contents contributers to Internet.}}

\author{ Chuyu Xiong \\
{\small Independent researcher, New York, USA} \\
{\small Email: chuyux99@gmail.com}
}

\begin{document}
\maketitle
\begin{abstract}
The number partition problem is a well-known problem, which is one of 21 Karp's NP-complete problems \cite{karp}. The partition function is a boolean function that is equivalent to the number partition problem with number range restricted. To fully understand the computational complexity of the number partition problem and the  partition function is quite important and hard. People speculate that we need new tools and methods \cite{aaronson} for such problem. In our recent research on the universal learning machine \cite{paper5, paper8}, we developed some tools, namely, fitting extremum, proper sampling set, boolean function with parameters (used in trial-and-error fashion). We found that these tools could be applied to the partition function. In this article, we discuss the set up of the partition function, properties of the partition function, and the tools to be used. This approach leads us to prove that the lower bound of the computational complexity of partition function, as well as the lower bound of the computational complexity of the number partition problem, is exponential to the size of problem. This implies: {\bf P} $\ne$ {\bf NP} \cite{cook}.   
\end{abstract}


{\it It is a great pleasant thing to learn and practice often!} \hspace{15pt}  \hfill--- Confucius \\
\vskip -15pt
{\it Simple can be harder than complex: You have to work hard to get your thinking clean to make it simple. ...... once you get there, you can move mountains.} \hspace{15pt}  \hfill--- Steve Jobs \\

\section{Introduction}
The number partition problem is one very famous problem \cite{wiki1}. It could be stated in a short sentence: given a set of natural numbers $\Omega$, can we divide $\Omega$ into two subsets $\Omega_1$ and $\Omega_2$ so that the sum of the numbers in $\Omega_1$ equals the sum of the numbers in $\Omega_2$?

\begin{exmp}[\bf Number Partition]
Given $\Omega = \{3, 1, 1, 2, 2, 1, 2, 2, 4\}$. If we set $\Omega_1 = \{1, 1, 1, 2, 4\}$ and $\Omega_2 = \{2, 3, 2, 2\}$, then $\sum \Omega_1 = \sum \Omega_2 = 9$. In this case, the partition problem has positive answer. Note, another partition is: $\Omega_1 = \{3, 1, 1, 2, 2\}$ and $\Omega_2 = \{1, 2, 2, 4\}$.

For set  $\Omega = \{3, 1, 2, 4, 1\}$, clearly the partition problem has negative answer, since the sum of $\Omega$ is odd. 

But for set $\Omega = \{3, 1, 2, 4, 12\}$, the partition problem has negative answer, even though the sum of  $\Omega$ is even.
\end{exmp}

Clearly, for a given set of integers $\Omega$, either $\Omega$ can be equally partitioned, or cannot. Thus, we have a boolean function, called as partition function. Clearly, the number partition problem is equivalent to the evaluation of the partition function. 

Detailed references of partition problem can be seen in \cite{wiki1}. The number partition problem is NP-complete \cite{karp, wiki3}. Yet, the number partition problem seems to be relatively easier than other NP-complete problems. It is often refereed as "the easiest hard problem" \cite{hayes} . There are many studies for this problem. Here, we are particularly interested in such question: what is the complexity of the partition function? Since a boolean function can be realized by a boolean circuit, thus, the equivalent question is: what is the complexity of the boolean circuit that realizes the partition function? 

Such a question is not easy. In order to address such question, we need new tool and new approach. Scott Aaronson said it quite elegantly: "find new, semantically-interesting ways to "hobble" the classes
of polynomial-time algorithms and polynomial-size circuits, besides the ways that have already been studied, such as restricted memory, restricted circuit depth, monotone gates only,
arithmetic operations only, and restricted families of algorithms..... Any such restriction that one discovers is
effectively a new slope that one can try to ascend up the $P\ne NP$ mountain." \cite{aaronson} We are highly encouraged by this line of thinking. But still, what is the new tools?

Our major research interests in recent years are in the universal learning machine and related problems \cite{paper2, paper5}. An universal learning machine is a machine that can learn any pattern from data without human intervention. In our setting, an universal learning machine has a conceiving space, and inside the conceiving space, there are a lot of X-forms. One X-form is nothing but a boolean function with subjectivity of the machine. Learning is actually to generate new X-forms and/or modify existing X-forms to meet data. Thus, we need to go deep on the relationship between data and boolean functions. Going along this path, we found a set of tools, namely, Fitting Extremum and Proper Sampling Set \cite{paper8}. Another set of tools are also related to the learning machine. We tried to study the subjectivity and the dynamic action of machine \cite{paper10}. In this process, we found Boolean Function with Parameters, which can be traced back to Kugel's Putnam-Gold machine \cite{kugel, kugel2}. In this article, we are going to try to use these tools on computational complexity of the partition function. 

This article is arranged in following way. In section 2, we will defined the partition vector and give detail definition of the partition function. One very useful lemma "Uniqueness of Partition Vector" is given, which reflects the deep nature about the partition problem. In section 3, we introduce the tool: Boolean function with Parameters. In section 4, we review Fitting Extremum (FE) and Proper Sampling Set (PSS), and 2 theorems (PSS implies circuit, circuit implies PSS). We also discuss some examples, which can help us to understand the partition function. In section 5, we use the tools on the partition function, which leads us to the conclusion. In section 6, we put forward some further thoughts that can help us to understand the methods and tools better.

\section{Partition Vector and Partition Function}
We need to have the exact definition of the partition function and to establish the relationship with the number partition problem. We will do so in several steps. First, we introduce the partition vector, which can help us to describe the number partition. Partition vector $p$ is a vector with components 1 or -1. This is natural: components 1 or -1 actually divide a group of numbers into 2 groups: one group with 1 and another group with -1, i.e. $p$ represents a way to partition.

\begin{definition}[\bf Partition Vectors]
A partition vector $p$ with length $N$ is a vector $p = (p_1, p_2, \ldots, p_N)$, with components $p_i = \pm 1, i=1, 2, \ldots, N$. For a group of natural numbers $\Omega = \{\alpha_1, \alpha_2, \ldots, \alpha_N\}$, the quantity $\sum_1^N p_i \alpha_i $, denoted as $<p, \Omega>$, represents the result of partition on $\Omega$ by $p$. Easy to see, $\Omega$ is equally partitioned by $p$ is equivalent to $<p, \Omega> = 0$. We call all partition vectors with length $N$ as Partition Vector Space, denote as $PV_N$, or just $PV$.
\end{definition} 

Note, if $p$ equally partitions a group of number $\Omega$, so does $-p$. Actually, $p$ and $-p$ represent exactly the same partition. Thus, we will only consider the partition vector with $p_1 = 1$. The simple examples below can help us to see how partition vectors are related to partition of a group of numbers.

\begin{exmp}[\bf Partition Vectors]
Given $\Omega = \{3, 1, 1, 2, 2, 1, 2, 2, 4\}$. Consider a partition vector $p = (1, -1, -1, 1, 1, -1, 1, 1, -1)$. Easy to see: $<p, \Omega> = \sum_1^9 p_i \alpha_i = 0$, i.e. $\Omega$ is equally partitioned by $p$. Another partition vector $p' = \{1, 1, 1, 1, 1, -1, -1, -1, -1\}$ also equally partitions $\Omega$, since $<p', \Omega> =0$. That is to say, for $\Omega$, there are more than one partition vector that equally partitions $\Omega$. 

Given $\Omega = \{3, 1, 2, 4, 1\}$, and partition vector $p = (1, -1, 1, -1, -1)$. $<p, \Omega> =\sum_1^5 p_i \alpha_i  = -1$. It is also easy to see that for any partition vector $p$, $<p, \Omega> \ne 0$. 

Given $\Omega = \{3, 2, 2, 1, 6\}$. Partition vector $p = (1, 1, 1, -1, -1)$ equally partitions $\Omega$. Moreover, $p$ is the only partition vector that can do so. Any other partition vector will not be able to equally partitions $\Omega$. In this case, we say $\Omega$ is uniquely equally partitioned by $p$, or $p$ is the unique partition vector for $\Omega$.  Note, when we say unique, we exclude the case: $p = (-1, -1, -1, 1, 1)$, since we only consider partition vector with $p_1 = 1$. 
\end{exmp}

In definition, the components of partition vector have values $\pm 1$. This is natural since we can directly apply multiplication and sum. Note, $\pm 1$ is binary value. But, normally, we use value 0 or 1 for binary values. Thus, we identify a usual boolean vector with a partition vector. The conversion between them is: $1 \leftrightarrow 1, -1 \leftrightarrow 0$. So, if $q \in \B^N$ is a normal boolean vector, it is equivalent to a partition vector $p \in PV_N$. So, we also say a boolean vector $q$ is one partition vector in this sense exactly. Thus, we can define $<q, \Omega> = <p, \Omega>$ for a group of number $\Omega$ and a boolean vector $q \in \B^N$. We are going to use this notation a lot. 

Note, when we say "a group of numbers", the numbers are natural number. But, we want to use the boolean function to study the problem. For this purpose, we need to restrict the range of numbers. For example, "numbers are less than 4", or "numbers are less than $2^N$", etc. Actually, such a range of number plays a crucial role in the number partition problem \cite{hayes}.  

Partition vector for a group of numbers is important and interesting. Let's consider some examples. 

\begin{exmp}[\bf Unique Partition Vector]
First, consider numbers in this range: integers less than 4. Take a partition vector, say, $p = \{1, 1, -1, -1, -1\}$. Easily see that $p$ equally partitions $\Omega =  \{3, 1, 1, 2, 1\}$. However, this $p$ is not unique for $\Omega$. Another partition vector $p' = \{1, -1, 1, -1, -1\}$ also equally partitions $\Omega$. 

Then consider the range of numbers: integers less than 8. Still consider the same partition vector $p = (1, 1, -1, -1, -1)$. Easy to see that $p$ equally partitions $\Omega =  \{6, 1, 3, 2, 2\}$. This time, $p$ is the unique partition vector of $\Omega$. 

At the above, first we have a partition vector $p$, then we look for a set of numbers (in certain range) $\Omega$, and to see if $p$ can equally partition $\Omega$, then to see if $p$ is unique. The range of numbers is important. 

We can also consider in opposite direction, i.e.  give a set of numbers $\Omega$, then look a partition vector $p$ so that  $<p, \Omega> = 0$, and if so, is $p$ unique? As one example, given $\Omega =  \{15, 8, 4, 2, 1\}$, look for $p$ with $<p, \Omega>=0$? Easy to see, $p = (1, -1, -1, -1, -1)$ is one, and $p$ is unique, i.e. for any other $p' \ne p $, $<p', \Omega> \ne 0$ 
\end{exmp}

The above simple examples actually raise an important question: given a partition vector $p$, can we find a set of numbers $\Omega$ so that $p$ equally partitions $\Omega$? For such $\Omega$, is $p$ unique? We can also look for the opposite question: given a set of numbers $\Omega$, can we find a partition vector $p$ that equally partitions $\Omega$? If so, is $p$ unique? We notice one important property: For some range of numbers, it is hard to find partition vector unique, while for some range of numbers, a partition vector is often unique. Such a property is deeply related to the number partition problem. In fact, \cite{mertens} discussed such property in some way. Following lemma will discuss these questions and property.

\begin{lemma}[\bf Uniqueness of Partition Vector]
For any $N>2$, for any partition vector $p$ with length $N$, we can find a group of numbers $\Omega = \{\alpha_1, \alpha_2, \ldots, \alpha_N\}$ with range $1 \le \alpha_i < 2^N$ so that $<p, \Omega>=0$, and $p$ is the unique partition vector to be so, i.e. for any other partition vector $p' \ne p$, must have $<p', \Omega> \ne 0$.  
\end{lemma} 
{\bf Proof: } We are going to do this: For a given partition vector $p$, find a group of numbers: $\Omega = \{\alpha_1, \alpha_2, \ldots, \alpha_N\}$ to satisfy the 2 conditions.

For convenience to talk, we introduce the term "sum of positive part" for $\Omega$ and $p$. $p$ is a partition vector, then $p_j, j=1, 2, \ldots, N$ is either 1 or -1. We call the sum: $\sum_{p_j = 1} p_j \alpha_j = \sum_{p_j = 1} \alpha_j $ as "sum of positive part", i.e. this sum is all $\alpha_j$ where $p_j=1$. Similarly, we have "sum of negative part": $\sum_{p_j = -1} \alpha_j $. Clearly, $<p, \Omega> =$ (sum of positive part) - (sum of negative part), and if $p$ equally partitions $\Omega$, then the "sum of positive part" equals the "sum of negative part". 

At first, we consider some special partition vectors that are like this: $p = (p_1, p_2, \ldots, p_N)$, with $p_1 = \ldots = p_J =1$, and $p_{J+1} = \ldots = p_N = -1$, where $1 \le J \le [N/2]$, $[N/2]$ is the integer part of $N/2$. That is to say, such partition vector has its first $J$ components equals $1$, and the rest of components equal $-1$. There are totally $[N/2]$ many such partition vectors. We consider these partition vectors one by one separately. 

For $J=1$, $p = (1, -1, -1, \ldots, -1)$, i.e. $p_1 =1$ and all other $p_j = -1$. For this partition vector $p$, we choose a group of numbers: $\Omega = \{2^N-2, 2^{N-1}, 2^{N-2}, \ldots, 2\}$, where $\alpha_1 = 2^N -2, \alpha_2=2^{N-1}$, etc. Clearly,  we have:
$$\text{sum of positive part} = \alpha_1 = 2^N -2$$
and 
$$\text{sum of negative part} = \sum_{j=2}^N \alpha_j = \sum_{j=2}^{N} 2^{N-j+1} = 2^{N} -2$$
So, $<p, \Omega> = \text{sum of positive part} - \text{sum of negative part} = 0$. For uniqueness, let's consider another partition vector $p' \ne p$. Note, $p'_1 = 1$, then there is at least one $k > 1$, so that $p'_k = 1$.  So, $<p', \Omega> = (\alpha_1 + \alpha_k) - (all \ other \ \alpha_j) > 0$. That is to say, the only possible partition vector to equally partition $\Omega$ is $p$.

For $J=2$, $p = (1, 1, -1, \ldots, -1)$, i.e. $p_1 = p_2 = 1$ and all other $p_j = -1$. For this partition vector $p$, we choose: $\Omega = \{\alpha_1, \alpha_2, \alpha_3, \alpha_4, \ldots, \alpha_N\}$, where $\alpha_1 = 2^N - 5, \alpha_2 = 1, \alpha_3 = 2^{N-1}, \alpha_4= 2^{N-2}, \ldots, \alpha_N =4$. Clearly, 
$$\text{sum of positive part} = \alpha_1 + \alpha_2 = 2^N -4$$
and 
$$\text{sum of negative part} = \sum_{j=3}^N \alpha_j = \sum_{j=3}^{N} 2^{N-j+2} = 2^{N} -4$$
So, $<p, \Omega> = \text{sum of positive part} - \text{sum of negative part} = 0$. For uniqueness, since the sum of $\Omega= \text{sum of positive part} + \text{sum of negative part} = 2(2^N -4)$, if $p'$ equally partitions $\Omega$, for $p'$ and $\Omega$, then, the sum of positive part (or negative part) must be $2^N-4$. Note, $p_1=1$ always, so it is easy to see that for this $\Omega$, the only possible partition vector $p'$ so that the sum of positive part equals $2^N-4$ is identical to $p$. So, if $p' \ne p$, must have $<p', \Omega> \ne 0$.

Generally, for any $J, 2 \le J \le [N/2]$, $p = (1, 1, \ldots, 1, -1, \ldots -1)$, i.e. $p_1 = \ldots = p_J =1$ and $p_{J+1} = \ldots = p_N = -1$. For this partition vector $p$, we choose: \\ 
$\Omega = \{\alpha_1, \alpha_2, \ldots, \alpha_J, \alpha_{J+1}, \ldots, \alpha_N\}$, 
where $\alpha_1 = 2^N - 2^J - 2^{J-1} +1, \alpha_2 = 2^{J-2}, \ldots, \alpha_J = 1, \alpha_{J+1} = 2^{N-1}, \alpha_{J+2} = 2^{N-2}, \ldots, \alpha_N = 2^{J}$. Clearly, 
$$\text{sum of positive part} = \sum_{j=1}^{J} \alpha_j = 2^N - 2^J - 2^{J-1} +1 + \sum_{j=2}^{J} 2^{J-2+j-2} = 2^N-2^J$$
and 
$$\text{sum of negative part} = \sum_{j=J+1}^N \alpha_j =  \sum_{j=J+1}^{N} 2^{N+J-j} = 2^N - 2^J$$
So, $<p, \Omega> =  \text{sum of positive part} - \text{sum of negative part} = 0$. For uniqueness, since the sum of $\Omega = \text{sum of positive part} + \text{sum of negative part} = 2(2^N - 2^J)$, if $p'$ equally partition $\Omega$, for $p'$ and $\Omega$, then, the sum of positive part (or negative part) must be $2^N-2^J$. Note,  $p_1=1$ always, so it is easy to see that for this $\Omega$, the only possible partition vector $p'$ so that the sum of positive part equals $2^N-2^J$ is identical to $p$. So, if $p' \ne p$, must have $<p', \Omega> \ne 0$.

Then we consider a partition vector $p$ with the number of 1's $\le [N/2]$. And, remember $p_1 =1$. We can permute the components of $p$ to get a partition vector $p'$, where $p'$ is one of those special partition vector discussed above. For example, $p = (1, -1, 1, -1, 1) \to p' = (1, 1, 1, -1, -1)$. So, for $p'$, we have $\Omega'$ as discussed above. Then, by reverse permutation on $\Omega'$, we get $\Omega$. Easy to see such $p$ and $\Omega$ satisfy: 1) $<p, \Omega>=0$, 2) $p$ is the unique partition vector to be so. 

Finally, we consider a general partition vector $p$, but the number of 1's could be bigger than $[N/2]$. In this case, we can consider $-p$, i.e. to change 1 to -1, and -1 to 1. We can get the $\Omega$ for this $p$, and with:  1) $<p, \Omega>=0$, 2) $p$ is the unique partition vector to be so. 
\hfill$\blacksquare$ \\

In the lemma,  for one partition vector $p$ we actually choose one $\Omega$ so that the 2 properties are hold. But, for a given $p$, there could be many $\Omega$ so that the 2 properties are hold. Specially, if the range of numbers are become bigger. Conversely, if the range of numbers are become smaller, we might not be able to choose $\Omega$ so that the 2 properties are hold. This is related to the discussions in \cite{hayes}. 

Here, we should make one note. In the above lemma, the range of numbers of $\Omega$ is $1 \le \alpha_i < 2^N$. If the range of numbers is $0 \le \alpha_i < 2^N$, if in $\Omega$, some $\alpha_i = 0$, then partition vector is could not unique anymore, $p_i$ can be either $1$ or $-1$. But, the lemma actually says: for a given  partition vector $p$, we can find a $\Omega$ with the range of numbers as $1 \le \alpha_i < 2^N$ so that $p$ is the unique partition vector for $\Omega$.  


Partition vector will make our descriptions on partition function much easier. However, if we want to form a boolean function, we also need to set a restriction on range for numbers, since we cannot use integer (which is infinite).

\begin{definition}[\bf Partition Function of Integers with Range]
For an integer $N>2$, and a natural number $M \ge 1$, define a function $Par$ as below:
$$
Par: \{\alpha_1, \alpha_2, \ldots, \alpha_N\} \to \B, \ \ 0 \le \alpha_i \le M 
$$
The value of $Par$ is: if there is one partition vector $p$ with length $N$ that equally partitions $\Omega = \{\alpha_1, \alpha_2, \ldots, \alpha_N\}$, i.e. $<p, \Omega>=0$, $Par(\Omega) = 1$, otherwise $Par(\Omega) = 0$. We call such function as the partition function with range $M$.   
\end{definition} 

The partition function captures the number partition problem, i.e. if the partition problem has positive answer, the partition function $Par = 1$, otherwise, $Par = 0$. Thus, to study the partition function is equivalent to study partition problem. But, here the partition problem is restricted: the number is not chosen from all integer, but only chosen from natural numbers not greater than $M$.  

The above partition function is well defined. However, it is not easy to calculate its value. But, at least, we can definitely calculate the value in this way: For a given $\Omega = \{\alpha_1, \alpha_2, \ldots, \alpha_N\}$,  we can traverse all partition vectors $p \in PV$, if for one $p$, $<p, \Omega>=0$, then stop, and $Par(\Omega) = 1$; if for any $p \in PV$, $<p, \Omega> \ne 0$, then $Par(\Omega) = 0$. Note, $PV$ has totally $2^{N-1} -1$ many partition vectors. So, on surface, the cost to calculate the value in this way is exponential to $N$.    

Here, we remark on the number of partition vectors in $PV$. As discussed above, we only consider such partition vectors whose first component is 1. And, we need to exclude the partition vector whose all components are 1 since this partition vector does not represent a partition at all. Thus, it is easy to see that the number of all possible partition vectors is $2^{N-1} -1$.

As above, the partition function is defined as a function on a set of numbers with range. We can modify this definition a little to define the partition function on bit array.

If we set $M$ in Def. 2.2 as $2^K-1$, i.e. consider $K$-bit integer, we then define the partition function of $K$-bits integers. We go further to turn a set of numbers into a bit array. There are several ways to turn an integer to bits, e.g. unary representation or binary representation. We will use the normal binary representation. That is to say, for a $K$-bit integer, its binary representation is a $K$ bit array. For a set of numbers, by concatenating bit arrays of all numbers in the set together, we will have a long bit array. Specifically, for a set of $K$-bit numbers $\Omega = \{\alpha_1, \alpha_2, \ldots, \alpha_N\}$, we have a bit array with length $KN$.  

For example, for the set of numbers: $\Omega =\{3, 1, 2, 7\}$, we have $N=4, K=3$, and $1=001, 2=010, 3 = 011, 7=111$, then $\Omega$ is represented by a bit array: $011001010111$. By this way, we can define the partition function of bit array.

\begin{definition}[\bf Partition Function of Bit Array]
For a integer $N>2$, and a integer $K \ge 1$, define a function $Par_{K,N}$ as below:
$$
Par_{K,N}: \B^{KN} \to \B, \ \ v = Par_{K,N}(x), 
$$
where $x \in \B^{KN}$ is a bit array with length $KN$, and $v \in \B$ is a binary value. The value of $Par_{K,N}$ is defined as: First, cut $x$ into a set of numbers of $K$-bits: $\Omega = \{\alpha_1, \alpha_2, \ldots, \alpha_N\}$, the cutting is from left to right; Second, if there is one partition vector $p \in PV_N$ that equally partitions $\Omega$, i.e. $<p, \Omega>=0$, $Par_{K,N}(x) = 1$, otherwise $Par_{K,N}(x)  = 0$. We call such function as the partition function of bit array with size $KN$.    
\end{definition} 

We also use $Par$ for $Par_{K,N}$ in the situation without confusing. When $K = N$, we use $Par_N$ for $Par_{N, N}$. 

\begin{exmp}[\bf Partition Function of Bit Array]
Consider a partition function with size $2 \times 6$. It is a boolean function $f: \B^{12} \to \B$. For any $x \in \B^{12}$, we can cut $x$ into 6 pieces, and each piece is a 2-bit integer (from 0 to 3).  For example, 
$$
x = 1101011010001  \  \rightarrow \{3, 1, 1, 2, 2, 1\}
$$
For this $x$, clearly, $Par_{2,6}(x) = 1$. Another example, 
$$
x = 1101011010000  \  \rightarrow \{3, 1, 1, 2, 2, 0\}
$$
For this $x$, should have $Par_{2,6}(x) = 0$. 
\end{exmp}

Clearly partition function with size $KN$ depends on $K$ and $N$. The combination of $K$ and $N$ will determine the property of the function. One special case is $K=1$. In this case, only 1 bit is used to represent integer. Thus, the only possible integer is 0 or 1. Easy to see, in this case, the partition function is reduced to a parity function. This case is simple but very useful.

As \cite{mertens} discussed, the complexity of partition function is determined by $K$ vs. $N$. Specially, when $K \ge N$, the partition function becomes complicated, hence interesting. Thus, we are going to consider $K = N$. So, for each $N$, the partition function $Par_N$ is a boolean function with dimension $N^2$. 

But, what about the dimension between square number $N^2$ and $(N+1)^2$? Consider a natural number $0 < L$, we want to have a boolean function $f_L: \B^L\to \B$  for each $L$, and when $L = N^2$, $f_L$ become $Par_{N,N}$ exactly. Here is the definition of such function.

\begin{definition}[\bf General Partition Function of Bit Array]
For a integer $L >2$, let $K = [\sqrt L]$, i.e. $K$ is the integer part of $\sqrt L$. And, let $N' = [L/K]$, i.e. the integer part of $L/K$. So, to define a boolean function $GPar_L$ as below:
$$
GPar_L: \B^L \to \B, \ \ v = GPar_L(b), 
$$
where $b \in \B^L$ is a bit array with length $L$, and $v \in \B$ is a binary value. Note, we have $KN' \le L < K(N'+1)$. There are 2 cases: \\
1. If $KN' = L$, then set $N=N'$, the value of $GPar_L$ is defined this way: First, cut $b$ into a set of numbers of $K$-bits: $\Omega = \{\alpha_1, \alpha_2, \ldots, \alpha_N\}$, if there is one partition vector $p \in PV_N$ that equally partitions $\Omega$, i.e. $<p, \Omega>=0$, $Par_L = 1$, otherwise $Par_L = 0$.  \\
2. If $KN' < L < K(N' +1)$, then set N = N'+1, the value of $GPar_L$ is defined as: First, cut $b$ into a set of numbers of $K$-bits: $\Omega = \{\alpha_1, \alpha_2, \ldots, \alpha_N\}$, Note, the cut is from left to right. The final number, i.e. $ \alpha_N$ has less than $K$-bits. But, it is still a number. If there is one partition vector $p \in PV_N$ that equally partitions $\Omega$, i.e. $<p, \Omega>=0$, $GPar_L= 1$, otherwise $GPar_L = 0$. \\
We call such function $GPar_L$ as general partition function of bit array with size $L$.    
\end{definition} 

It is easy to see: if $L$ is a square number, i.e. $L = N^2$, then $GPar_L = Par_{N,N} = Par_N$. 

\begin{exmp}[\bf Examples of General Partition Function of Bit Array]
One example: Consider $L = 12$, so, $K = [\sqrt L]  =3$, $N = [L/K] = 4$, and $L=KN$. Thus, we are going to cut bit array into 4 pieces, each piece 3-bits. One case: 
$$
x = 110101101000  \  \rightarrow \{6, 5, 5, 0\}
$$
For this $x$, clearly, $GPar_{12}(x) = 0$. 
Another example: Consider $L = 13$, so, $K = [\sqrt L]  =3$, $N = [L/K] = 4$, and $L=KN + 1$. Thus, we are going to cut bit array into 5 pieces, first 4 pieces are 3-bits, last piece is 1-bit. One case:
$$
x = 1001011010111  \  \rightarrow \{4, 5, 5, 3, 1\}
$$
For this $x$, easy to see $GPar_{13}(x) = 1$. 
\hfill$\blacksquare$
\end{exmp}

The computational complexity of the boolean functions $GPar_L: \B^L \to \B, \  L = 2, 3, 4, \ldots$ is the topic we concern. Particularly, when $L = N^2, N = 2, 3, \ldots$, we have $Par_N$, which is equivalent to the number partition problem with range restricted to $2^N-1$.

\section{Boolean Function with Parameters}
In this section, we are going to introduce one tool, i.e. boolean function with parameters. We realized this tool when we studied subjectivity and dynamic action of machine \cite{paper10}. In the research, we noticed Kugel's Putnam-Gold machine \cite{kugel, kugel2}. Kugel thought a Turing machine can be used in trial-and-error fashion and he argued that by such a way, a Turing machine could do much more. Such a thought gives us an inspiration: If we introduce boolean function with parameters and use it in trial-and-error fashion, we can have a new way to write boolean function, and such way is quite powerful to express some functions otherwise hard to express. So far, we have not seen any reference for such concept, i.e. boolean function with parameters to be used in such a fashion. This tool is quite useful, so we write this section to explain it. First, consider one very simple example.

\begin{exmp}[\bf A Simple Case]
$f(x_1, x_2) = x_1 \land x_2$ is a simple boolean function on $\B^2$ to $\B$. It has no parameters. Based on it, we introduce one boolean function with parameters: $f(x_1, x_2, s, t) = x_1{}^s\!\land^t x_2$. Here, parameter $s$ is one boolean variable behaving like a switch: if $s=1$, do nothing, if $s=0$, put into a $\neg$. Same for $t$. Thus, $f(1, 1, 0, 1) = 1 \neg \land 1 = 0 \land 1 = 0$, and $f(0, 1, 0, 1) = 0 \neg \land 1 = 1 \land 1= 1$. That is to say, with different parameters, the function actually is a different boolean function. In fact, $f(x_1, x_2, 1, 0) = x_1 \land \neg x_2$, and $f(x_1, x_2, 0, 1) = \neg x_1 \land x_2$. 
\end{exmp}

The definition of a boolean function with parameters is below.

\begin{definition}[\bf Boolean Function with Parameters]
A boolean function $\varphi: \B^N \times \B^J \to \B$, is called as a boolean function on $\B^N$ with $J$ binary parameters. We often write such function as $\varphi(x, p): \B^N \to \B$, where $x \in \B^N, p \in \B^J$. 
\end{definition} 

The boolean function with parameters can be used in trial-and-error fashion. Suppose $\varphi$ is a boolean function with parameters, and we have a list of parameter vectors: $P = \{p_1, p_2, \ldots, p_K\}$, we can conduct a trial-and-error process: for a given $x$, first try $\varphi$ with $p_1$, if $\varphi(x, p_1) = 1$, it is good, we stop trial and $f(x) =1$; if $\varphi(x, p_1) = 0$, trial fails, we continue to try $\varphi$ with $p_2$, and do the same as for $p_1$; if at some $p_k$, $\varphi(x, p_k) = 1$, trial successes at $p_k$ and $f(x) = 1$; if for all parameter vector $p_k \in P$, trial fails (i.e. $\varphi(x, p_k) = 0$), then $f(x) = 0$. We can see, by this way, we get a new boolean function from $\varphi$ and $P$. This is a quite essential process. We formally define it as following. 

\begin{definition}[\bf Trial with Parameters]
Suppose $\varphi: \B^N\times \B^J \to \B$ is a boolean function with $J$ binary parameters, and $P = \{p_1, p_2, \ldots, p_K\}$ is a list of parameter vectors (each $p_k$ is a $J$-dim binary vector), we can form a boolean function $f: \B^N \to \B$ in this way: for a given input $x \in \B^N$, if there is one parameter $p_k$ so that $\varphi(x, p_k) = 1$, then $f(x) = 1$; otherwise, i.e. for all parameter vector $p_k \in P$, $\varphi(x, p_k) = 0$, $f(x) =0$. Such function $f$ is called as $\varphi$ trial with $P$, and we denote as: $f = \varphi \odot P = \varphi \odot \{p_1, p_2, \ldots, p_K\}$. 
\end{definition} 
 
Here is a note for the symbol $\odot$. It is used to represent the operator of a boolean function with parameters trial with a parameter list. This operator is similar to "shooting target", while the symbol looks like a simplified target. It is quite intuitive.

\begin{exmp}[\bf Simple Case of Trial with Parameters]
Here is one very simple example. As in Example above, $\varphi(x_1, x_2, s, t) = x_1{}^s\!\land^t x_2$ is a boolean function on $\B^2$ with 2 parameters. Consider a list of parameters: $S= \{(1,0), (0,1)\}$. It is quite easy to see that $\varphi$ trial with $S$ equals parity function, i.e. $\varphi \odot S = x_1 \oplus x_2$. 
\end{exmp}

In fact, the boolean function with parameters is very much targeting to the partition function. It is very easy to see that using trial-and-error fashion to express the partition function is very natural. First, we define a boolean function with parameter, which actually represent the partition by one particular partition vector. Exactly, we define a boolean function with parameters $\varphi$ by following equation: 
\begin{equation}
\varphi: \B^{KN}\times \B^N \to \B, \ \ \varphi(x, p) = \begin{cases} 
      1 & \text{if $<p, \Omega_x> =0$} \\
      0 & \text{if $<p, \Omega_x> \ne 0$} 
\end{cases}
\end{equation}
where $x \in \B^{KN}$ is a $KN$-dim bitarray, $p \in \B^N$ is partition vector ($N$-dim), $\Omega_x$ is the set of numbers cut from $x$, $\Omega_x$ has $N$ $K$-bit numbers,  and $<p, \Omega_x>$ is the sum defined Def. 2.1 (note the conversion: $1 \leftrightarrow 1, -1 \leftrightarrow 0$). Clearly, the function $\varphi$ is a boolean function with parameters. The meaning of $\varphi$ is very clear: if $x$ can be equally partitioned by $p$, $\varphi(x, p) =1$, otherwise $\varphi(x, p) =0$. 

Using this notation, we can restate the Uniqueness of Partition Vector lemma again as below. This will be quite useful.

\begin{lemma}[\bf Uniqueness of Partition Vector]
For any $N>2$, for any partition vector $p$ with length $N$, we can find a boolean vector $x \in \B^{N^2}$ so that $\varphi(x, p)=1$, and $p$ is the unique partition vector to be so, i.e. for any other partition vector $p' \ne p$, must have $\varphi(x, p')=0$.  
\end{lemma} 

How to use $\varphi$? We are going to use it by trial-and-error fashion. That is to say, if $P$ is a set of partition vectors, we can define a boolean function $f = \varphi \odot P$. The purpose to do so is clearly shown in next lemma.

\begin{lemma}[\bf Partition Function expressed by trial-and-error]
Partition function of bitarray $Par_{K,N}: \B^{KN} \to \B$ can be expressed by trial-and-error fashion: $Par(x) = \varphi(x, p) \odot PV$, where $\varphi$ is the boolean function with parameters defined in Eq. 1, and $PV_N = PV$ is the partition vector space. 
\end{lemma} 

The proof of the lemma is directly from the definition of $Par_{K,N}$ and $\varphi$. Thus, we can see: partition function can be very easily expressed by boolean function with parameters and trial with parameters. Compare to the definition of partition function in Def. 2.3, the way to express it in trial with parameters are very natural and much easier to handle.

In above defintion, $\varphi$ trial with the whole partition vector space $PV$, and we get partition function. But, how about trial with only some partition vectors? In such case, we get the sub-partition function. Definition is below.

\begin{definition}[\bf Sub-Partition Function]
Suppose $P \subset PV_N$, i.e. $P$ is a set of partition vectors, a sub-partition function on $P$ is a boolean function $SPar_{K,N,P}: \B^{KN} \to \B$, $SPar_{K,N,P}(x) = \varphi(x, p) \odot P$, where $\varphi$ is the boolean function with parameters defined in Eq. 1. 
\end{definition} 

Note, if $P = PV$, then $SPar_{K,N,P} = Par_{K,N}$. 

A boolean function can be expressed by a boolean circuit. We want to show, similarly, a boolean function with parameters can be expressed by a boolean circuit with parameters. 

First, we need to consider how to build parameters into circuit. There are several ways. One way is to consider: a boolean circuit with switches is constructed by nodes of $^s\land^t$ and $^s\lor^t$, where $s$ and $t$ are switches that can take the position of "pass" or "negation" according to the value of parameter $s$ or $t$: if $s=1$, it is "pass"; if $s=0$, it is "negation"; similarly for $t$. In Example 3.1, we have seen such boolean circuit with parameters. Another way is to consider: in the circuit, add some "constant nodes", which takes fixed value (but, we can switch the value according to parameter value). For example, $C$ is a boolean circuit. We add one constant node $o$ into, and form a new circuit: $C'_p=C \land o$. This new circuit will perform differently for different value of $o$. If $o=1$, $C'_p$ becomes $C$, if $o=0, C'_p=0$. This circuit $C'_p$ is a boolean circuit with parameters. 

It is easy to show that the 2 ways are equivalent. But we are not going to discuss it here. In below lemma, we are going to use the second way. 

\begin{lemma}[\bf Boolean Circuit with Parameters]
Any boolean function with parameters $\varphi$ can be expressed by a boolean circuit with parameters $C$.
\end{lemma} 
{\bf Proof:} Consider a boolean function with parameters $\varphi: \B^N \times \B^J \to \B$. We are going to show that we can find a boolean circuit with parameters to express $\varphi$. 

For any given parameters $p \in \B^J$, $\varphi(., p)$ is a boolean function on $\B^N$, then we can find a circuit $C_p$ expressing $\varphi(., p)$. Consider all such circuits: $\{C_p \ | \  p \in B^J\}$. There are totally $L = 2^J$ many such circuits. We are going to make a circuit with parameters from these circuits $\{C_p\}$.

Then, we build a circuit with parameters $O_p^q$ like this: $p \in \B^J$ is the parameter, $q \in \B^J$ is a structure indicator that indicates the specific structure of this circuit, circuit is $O_p^q = ^so_1 \land ^so_2 \ldots \land ^so_J$, where $o_j$ is a constant node that takes value $p_j$, $s$ are switches that takes value as: if $q_j=1$, $s$ is pass, if $q_j=0$, $s$ is negation. The circuit with parameters $O_p^q$ has this property: if parameter vector equals structure indicator, i.e. $p=q$, $O_p^q = 1$, else, i.e. $p \ne q$, $O_p^q = 0$. 

Using $O_p^q$ and $C_p$, we can get a circuit with parameters as following: 
$$
V(p) = \bigvee\limits_{q \in \B^J} (C_q \land O_p^q) 
$$
We can clearly see that $V(p)$ is a boolean circuit with parameters expressing $\varphi$, since $\forall x \in \B^N, \forall p \in \B^J$, we have $V(p)(x) = C_p(x)$.     
\hfill$\blacksquare$ \\

In the above proof, we only consider how to express $\varphi$ and did not consider any other factors. In fact, the above circuit $V(p)$ often is not an efficient one. Quite often, we can make a much more efficient circuit.

Finally, in Eq. 1, we defined a boolean function with parameters, what is the circuit with parameters expressing this function? This is highly related to the partition function. We will see it in next section.

\section{Fitting Extremum and Proper Sampling Set}
Now, we discuss another set of tools. In recent years, we have been trying to see why and how a machine can learn from data without human intervention (so called mechanical learning, since without human factor means following a mechanical rule) \cite{paper2, paper5}. An universal learning machine is a theoretical model for such purpose, which contains conceiving space, and in conceiving space, there are many X-forms. Actually, if without considering subjectivity of machine, a X-form is one boolean circuit (expressing a boolean function). In the process to understand universal learning machine, we found that mechanical learning can be achieved by conducting fitting extremum \cite{paper8}. Fitting Extremum is, very briefly say, if a learning machine keeps looking for a boolean circuit that fits with data with minimal number of nodes ($\land, \lor$ nodes), then eventually, that boolean circuit will express the boolean function desired to learn. Such a learning is pure mechanical, no human intervention is necessary. Just follow this rule, the learning can be achieved if there is enough data, no matter what is the learning target. The data sufficient for this purpose is Proper Sampling Set. This result so far is only theoretical, but it reveals the nature of learning. We are currently conduct research to push it to applications. 

During the research process, we eventually realized Fitting Extremum (FE) and Proper Sampling Set (PSS) forming good tools for computational complexity, specially for the partition function. Moreover, FE and PSS are working well with the boolean function with parameters and trial-and-error fashion together. 

In this section, we will review FE and PSS. We just give necessary definitions and results. For details, please see \cite{paper8}.

Below, we are considering to learn a boolean function $f: \B^N \to \B$ from data.  

\begin{definition}[\bf Sampling Set]
A sampling set $S$ is one subset of $\B^N$, i.e. $S \subset \B^N$. We also just say sampling. Moreover, over one sampling set, there are assigned values:
$$
Sv = \{ [x, b] \ | \ x \in S, \ b = \text{0 or 1}  \}
$$
We call such set $Sv$ as sampling set with assigned values, or sampling with values, or just sampling. For a boolean function $f: \B^N \to \B$,  we can have the sampling set with values of $f$ (or just sampling for $f$): 
$$
Sv = \{ [x, f(x)] \ | \ x \in S \}
$$
\end{definition}

With sampling, learning is to find a boolean function that fits with sampling. Actually, learning is looking for a boolean circuit that fits with sampling. But, for one sampling, there could be many possible such boolean functions. Note, the set of all boolean functions on $\B^N$ has huge size: $2^{2^N}$. Moreover, if we denote all boolean circuits on $\B^N$ as $\C$, the size of $\C$ is even bigger than $2^{2^N}$, since a boolean function could have many boolean circuits expressing it. So, what do we do? 

Fitting Extremum is to look for a boolean circuit that has least number of nodes while fitting with sampling. So, we need to consider the number of nodes of the boolean circuit. For a boolean circuit $C$, the number of nodes of $C$ $d(C)$ by our definition \cite{paper8}, is slightly different from the number of nodes of $C$ $d'(C)$ usually used, since we do not count negation as one node. But, the two number of nodes have a simple relationship:  $d(C) < d'(C)$ and $d'(C) < 2d(C)$. Thus, there will no trouble to use $d(C)$ to replace $d'(C)$ in our later discussions.   

\begin{definition}[\bf Fitting Extremum]
For a sampling set with values $Sv$, we define one extremum problem as following:
$$
\text{Min:}\ d(C), \ C \in \C  \ \& \  \forall [x, b] \in Sv \ C(x) = b 
$$
We call this problem as fitting extremum on $Sv$.
\end{definition} 

In the definition of fitting extremum, we give a sampling set with values. But, what if we give a subset of $\B^N$ and a boolean function? This sure will define a fitting extremum as well.  

\begin{definition}[\bf Fitting Extremum of a Boolean Function]
For one boolean function $f: \B^N \to \B$, and for a sampling set $S \subset \B^N$, we define one extremum problem as following:
$$
\text{Min:}\ d(C), \ C \in \C  \ \& \  \forall x \in S \ C(x) = f(x) 
$$
We call this problem as fitting extremum on $S$ and $f$.
\end{definition}   

Such a circuit $C$ is called as the circuit generated by FE on sampling $S$ and function $f$. That is to say, given a sampling and a boolean function, we can generate a circuit from them by FE. We will say FE on $S$ to fit $f$, or FE on $S$ by fitting $f$ to gets $C$, etc. 

Note, for a given $S$, the circuit generated from FE on $S$ to fit $f$ is often not $f$. Could be totally different than $f$. However, when $S$ satisfies certain condition, the circuit will express $f$.

\begin{definition}[\bf Proper Sampling Set]
For a given boolean function $f: \B^N \to \B$, and for a sampling set $S \subset \B^N$, if FE on $S$ and $f$ generates a boolean circuit $C$, i.e. $C$ fits $f$ on $S$, and $d(C)$ reaches minimum, and if $C$ expresses $f$ exactly, i.e. $\forall x \in \B^N, C(x) = f(x)$, we say $S$ is a proper sampling set of $f$, or just proper sampling. 
\end{definition} 

We will use FE to stand for fitting extremum and PSS for proper sampling set. In another words, when $S$ is a PSS of $f$, the boolean circuit generated by FE on $S$ to fit $f$ will express $f$. This is one crucial property. Again, if $S$ is a PSS of $f$, FE on $S$ to fit $f$ will get a circuit expressing $f$.


\begin{lemma}[\bf Existence of PSS]
For any boolean function $f$, there is some subset $S \subset \B^N$ so that $S$ is PSS of $f$. 
\end{lemma}

The lemma says that for any boolean function $f$, PSS of $f$ always exists. The proof is very easy to see. The trivial (and worst) case is that PSS equals the whole boolean space $\B^N$. We can think in this way: give a sampling $S$, if $S$ is not PSS, we can add more elements into $S$. Since $\B^N$ is finite, eventually, $S$ will become PSS. Of course, we do not want the whole space, if possible. So, we need to consider a PSS as small as possible. 
 
\begin{definition}[\bf Minimal PSS]
For a given boolean function $f: \B^N \to \B$, if a sampling $S \subset \B^N$ is a proper sampling set, and $|S|$ reaches the minimum, we call such a sampling set as minimal proper sampling set. 
\end{definition} 

We use mPSS to stand for minimal PSS. If $S$ is a mPSS of a boolean function $f$, then for any sampling $S'$, if $|S'| < |S|$, $S'$ could not be a PSS of $f$. Thus, mPSS of $f$ indeed describes one important property of $f$. It is easy to see that for any boolean function $f: \B^N \to \B$, mPSS indeed exists.

\begin{lemma}[\bf mPSS]
For a boolean function $f: \B^N \to \B$, mPSS exists, i.e. there is a sampling $S$, $S$ is PSS of $f$ and for any sampling $S'$, if $|S'| < |S|$, $S'$ could not be a PSS of $f$.
\end{lemma}
{\bf Proof:} We know that PSS of $f$ indeed exists. But, all possible PSS of $f$ form a finite set. We choose the sampling in this set with smallest size, it will be a mPSS of $f$.  
\hfill$\blacksquare$

For one $f$, there could be more than one mPSS, i.e. might have such a situation: $S_1$ and $S_2$ are both mPSS of $f$, and $S_1 \ne S_2$. But, the size of all mPSS of $f$ is same. Thus, the size of mPSS gives us one important property of $f$.

FE and PSS are deeply related to learning. Particularly, the size of sampling set is deeply related to complexity of boolean circuit. Following 2 theorems tell such a relationship.

\begin{theorem}[\bf PSS implies Circuit]
If $f$ is a boolean function $f: \B^N \to \B$, and $S \subset \B^N$ is a PSS for $f$, and $|S|$ is the size of PSS, then there is a circuit $C$ expressing $f$ and $d(C) < N|S|$. 
\end{theorem}

Opposite direction is also true: If we have a circuit, we can construct a PSS from it. 

\begin{theorem}[\bf Circuit implies PSS]
If $f$ is a boolean function $f: \B^N \to \B$, and $C$ is a boolean circuit expressing $f$, then there is a PSS for $f$, and size of PSS is bounded by $3d(C)$. 
\end{theorem}

The 2 theorems tell us that for a boolean function $f$, if we have a PSS of $f$, we can construct a circuit to express $f$ and the size of circuit ($d(C)$) is proportional to the size of sampling. And, reversely, if there is one circuit expressing $f$, then we can find a PSS by using circuit, and the size of sampling is controlled by size of circuit. Since the size of circuit ($d(C)$) is one good measure of computational complexity of $f$, so is the size of PSS. This is a very important property, which means that we can examine the complexity of $f$ by examining the learning process and PSS of $f$. Particularly, "circuit implies PSS" gives a lower bound of circuit by the size of mPSS. This is the tool we are going to use. We have not so far seen such a tool in any literature. Thus, we think this is a new tool for computational complexity.
\footnote{When we worked on FE and PSS in 2019, we did not know there were works on "partially defined Boolean function" (pdBf) \cite{crama}. Our process of learning can be thought as a series of pdBf that is expanding. However, FE is much different than pdBf. pdBf did not seek fitting extremum, and did not relate to learning. }
\footnote{Conceptually, PSS is related to sampling complexity \cite{wiki4}. But, so far we have not seen any work on sampling complexity in the direction to seek fitting extremum, hence related to complexity of circuit.}

\subsection*{Examples of Boolean Functions and Circuits}
We are going to see some examples of boolean functions, their PSSs and their circuits. These examples are simple boolean functions, but highly related to the partition function and are very useful. 

\subsubsection*{Simplest Addition and Subtraction}
We consider the simplest addition: $z=x+y$, where $x=x_1x_0, y=y_1y_0$ are 2 bits integers, $z=z_2z_1z_0$ is a 3 bits integer. For example, $x=01, y=10, z=x+y=011; \  x=11, y=01, z=x+y=100$. Easy to see such addition can be completely described by following 3 boolean functions $z_0, z_1, z_2$. 
\begin{equation}
z_0 = x_0 \oplus y_0,  \ \ t_0 = x_0 \land y_0, \ \
z_1 = (x_1 \oplus y_1) \oplus t_0, \ \ t_1 = (x_1 \land y_1) \lor (x_1 \land t_0) \lor (y_1 \land t_0), \ \
z_2 = t_1
\end{equation}
Here, $t_0, t_1$ are 2 adjunct functions for carry-over value for $x_0+y_0$ and $x_1+y_1+t_0$.  All these 5 boolean functions are on $\B^4$. Note, elements in $\B^4$ is formed in this way: $x_1x_0y_1y_0$. In another words, an element in $\B^4$ is cut into $x, y$, then do the addition.

It is very interesting to see the PSSs and circuits of these boolean functions. First consider $z_0$ for addition, which is a boolean function ($\oplus$) on $\B^4$. Pick a sampling set in $\B^4$, say $S=\{0000, 0100, 0001, 0101\}$. We can easily see such $S$ is a mPSS of $z_0$. It means: FE on $S$ to fit $z_0$ will get a circuit $C$, then $C$ must express $z_0$. Should note, there are more mPSS. For example, $S=\{1010, 0100, 0001, 0101\}$ is another.  Note, $|S| = 4$.  

Second, consider $t_0$ for addition. This time, we can pick up a sampling in $\B^4$ like this: $S=\{0100, 0001, 0101\}$. This is a mPSS for $t_0$. Should note, any mPSS for $z_0$ is a PSS for $t_0$. 

Then, consider $z_1$ for addition. $z_1$ is formed by 2 branches: $x_1 \oplus y_1$ and $t_0$. The sampling for $x_1 \oplus y_1$ could be: $S=\{0000, 1000, 0010, 1010\}$. The sampling for $t_0$ could be: $S'=\{0000, 0100, 0001, 0101\}$. However, $S\cup S'$ could not be PSS for $z_1$. We can add $S''=\{0000, 0100, 1000, 1100\}$ to make sure to capture the second $\oplus$. $\hat{S} = S\cup S'\cup S'' = \{0000, 1000, 0010, 1010, 0100, 0001, 0101, 1100\}$ is mPSS of $z_1$. Note $|\hat{S}| = 8$.

Then, consider $t_1$ for addition. We will add more to $\hat{S}$. This set $S''' =  \{0101, 1100\}$ is good for capture the $\lor$ and $\land$ in $t_1$. So, the set $\bar{S} = \hat{S} \cup S'''$ is a mPSS of $t_1$. Note, $|\bar{S}| = 10$. $z_2 = t_1$, so $\bar{S}$ is PSS of $z_2$. 

We can easily see: a PSS for $z_2$ is also a PSS for $z_0$ and $z_1$, so, $\bar{S}$ is a PSS for all those 5 boolean functions. This sampling set $\bar{S}$ is sufficient to describe the addition.  

Then, we consider the simplest subtraction: $z = x-y$, where $x=x_1x_0, y=y_1y_0$ are 2 bits integers, $z = z_2z_1z_0$ is a 3-bit integer with first digit for sign (0 for positive, 1 for negative). For example, $x=01, y=10, z=x-y=101; x=11, y=01, z=x-y=010$. 

We can use addition operation to get result of subtraction. It is like this: For $x-y$, if $x < y$,  then do additions: $z=\hat{x}+1+y$, then $z_2z_1z_0$ is the result of subtraction, where $z_2=1$, $z_1, z_0$ are from the addition; if $x \ge y$, then do addition: $z=x+\hat{y}+1$, then $z_2z_1z_0$ is the result of subtraction, where $z_2=0$, $z_1, z_0$ are from the addition. Here, $\hat{1} =0, \hat{0} = 1$, $\hat{x_1x_0} = \hat{x_1}\hat{x_0}$, etc. For example, $x=01, y=11$, for $x-y$, do addition $z=10+11+01$, then $110$ is the result of subtraction. Another example, $x=11, y=10$, for $x-y$, do addition $z=11+01+01$, then $001$ is the result of subtraction. 

When we do subtraction, we need to have one function to judge which number is greater, i.e. this boolean function $g(x): \B^4 \to \B$, value of $g(x)$ are: for $x=x_3x_2x_1x_0 \in \B^4$, let $x=x_3x_2, y=x_1x_0$, if $x \ge y$,  then $g(x)=1$, if $x < y$, then $g(x)=0$. We also need another boolean function to judge the equality: $v(x): \B^4 \to \B$, value of $v(x)$ for $x=x_3x_2x_1x_0 \in \B^4$ are: if $z=x_3x_2-x_1x_0=0$, then $v(x)=1$, else, i.e. $z=x_3x_2-x_1x_0 \ne 0$, then $v(x)=0$. 

PSS for $g$ and $v$ are very useful. These PSSs together with PSSs of additions will fully describe subtraction. 
 
These boolean functions above are very simple. But, they are quite illustrative and useful.

\subsubsection*{Addition and Subtraction of $K$-bits Numbers}
Above, we considered addition, subtraction and partition function for 2-bits integers, as well as related circuits and PSSs. This helps us to understand better. Now, we consider addition and subtraction for $K$-bits integer. First, consider addition $z=x + y$, where $x=x_{K-1}\ldots x_1x_0, y=y_{K-1}\ldots y_1y_0$ are $K$-bits integers, $z=z_K\ldots z_1z_0$ is a $(K+1)$-bits integer. For example, $x=1001, y=1010, z=x+y=10011$. Easy to see such addition can be described by following $K+1$ boolean functions for $z_0, z_1, \ldots, z_K$ (and $K$ adjunct boolean functions):
\begin{equation}
\begin{split}
& z_0 = x_0 \oplus y_0,  \ \ t_0 = x_0 \land y_0, \\
& z_1 = (x_1 \oplus y_1) \oplus t_0, \ \ t_1 = (x_1 \land y_1) \lor (x_1 \land t_0) \lor (y_1 \land t_0), \\
& \ldots \ldots \ , \\
& z_{K-1} =  (x_{K-1} \oplus y_{K-1}) \oplus t_{K-2}, \ \ t_{K-1} = (x_{K-1} \land y_{K-1}) \lor (x_{K-1} \land t_{K-2}) \lor (y_{K-1} \land t_{K-2}), \\
& z_K = t_{K-1}
\end{split}
\end{equation}
Here, $t_0, t_1, \ldots, t_{K-1}$ are $K$ adjunct functions for carry-over value for $x_0+y_0, x_1+y_1$ etc.  All these $2K+1$ boolean functions are on $\B^{2K}$. Note, elements in $\B^{2K}$ is formed in this way: $x_{K-1}\ldots x_1x_0y_{K-1}\ldots y_1y_0$. In another words, an element in $\B^{2K}$ is cut into $x, y$, then do the addition. 

We can use addition operation to get result of subtraction. It is like this: For $x-y$, if $x < y$,  then do additions: $z=\hat{x}+1+y$, then $z_K z_{K-1}\ldots z_1z_0$ is the result of subtraction, where $z_K=1$, $z_{K-1}, \ldots, z1, z_0$ are from the addition; if $x \ge y$, then do addition: $z=x+\hat{y}+1$, then $z_K z_{K-1}\ldots z_1z_0$ is the result of subtraction, where $z_K=0$, $z_{K-1}, \ldots,z1, z_0$ are from the addition. Here, $\hat{1} =0, \hat{0} = 1$, $\hat{x_{K-1}\ldots x_1x_0} = \hat{x_{K-1}}\ldots \hat{x_1}\hat{x_0}$, etc. 

For example: $K=5, x=10101, y=11010$, since $x < y$, do addition $z = \hat{x} + y +1$, then $z=x-y=100101$. Another example: $K=4, x=1101, y=1010$, since $x > y$, do addition $z = x + \hat{y} +1$, then $z=x-y=00011$. 

In order to do subtraction, we also need this boolean function: $g(x, y): \B^K\times \B^K \to \B$, value of $g(x, y)$ are: if $x \ge y$,  then $g(x, y)=1$, if $x < y$, then $g(x, y)=0$, where $x_{K-1}\ldots x_1x_0$ is the binary representation of $x$, and $y_{K-1}\ldots y_1y_0$ is the binary representation of $y$.   

For these boolean functions in addition, it is very interesting to see their circuits and PSSs. One immediate observation is: $z_1$ with $t_0$ are almost same as $z_2$ with $t_1$, $\ldots$, and almost same as $z_{K-1}$ with $t_{K-2}$. They have exactly same structure, the only difference is the index. This is no surprise. Addition and subtraction are linear to $K$ (bits of integer). Such a property will make to find circuits and PSSs much easier. 

We already know the PSS for $z_1$ with $t_0$ (above $\hat{S}$). The PSS for $z_2$ with $t_1$ is just linearly expansion from $\hat{S}$. And, this linear expansion will continue to $z_{K-1}$ with $t_{K-2}$, and to $z_{K}$ with $t_{K-1}$. Here, we will not write down exactly PSSs for them. But, it is very clear that the PSS of these boolean function has size proportional to $K$. So does mPSS $S$. Such $S$ will be sufficient to describe all boolean functions for addition and subtraction, so sampling $S$ is sufficient to describe the addition and subtraction. Notice, in $\B^{2K}$, there are totally $2^{2K}$ many elements. So, relatively, $S$ is a very small subset in $\B^{2K}$.

\subsubsection*{Put Together with One Partition Vector}
We have seen addition/subtraction. We then consider how to computer the function in Eq. 1. For a given partition vector $p$, the computation of $\varphi(x, p)$ is done by $N-1$ additions/subtractions on $K$-bits integers, where $p_j$ determines to do addition or subtraction at $j$-th place. So, for a given $p$, $\varphi(., p)$ is a boolean function on $\B^{KN}$.  We want to consider the PSS and circuit for this boolean function. 

Consider $x \in \B^{KN}$, and $p \in PV_N$. As in Eq. 1, we cut $x$ into a group of numbers: $\Omega_x = \{\alpha_1, \alpha_2, \ldots, \alpha_N\}$, each number is a $K$-bits number. We first compute $z =<p, \Omega_x>$. Here, $z$ is a $K+N-1$ bits integer. If we have $z =z_L\ldots z_2z_1$, where $L=K+N-1$, then $\varphi(x, p)$ is: 
$$
\varphi(x,p) =  \bigwedge\limits_{j=1,2, \ldots, L} \neg z_j
$$
Note, $z = <p, \Omega_x>$ is actually formed by $N-1$ addition/subtraction. We have shown above, each addition/subtraction can be fully described by at most $2(K+N+1)$ boolean functions, and all these boolean functions can be fully described by a PSS $S$, whose size is linear to $2(N-1)(K+N+1)$, or linear to $NK+N^2$. So, the size of circuit expressing $\varphi(.,p)$ is also linear to $NK+N^2$. Since we often consider $K \ge N$, the size of circuit will be proportional to $KN$.

\subsubsection*{Boolean Function with Parameters}
No, we consider the circuit with parameters expressing the boolean function with parameters in Eq. 1. Suppose $C_p$ is the circuit expressing $\varphi(., p)$, the method in lemma "Boolean Circuit with Parameters" tells us how to establish a circuit with parameters expressing $\varphi(x, p)$. 

To do so, let's see a family of circuit with parameters $O_p^q$: $p \in \B^J$ is the parameter, $q \in \B^J$ is a structure indicator that indicates the specific structure of this circuit, circuit is $O_p^q = ^so_1 \land ^so_2 \ldots \land ^so_J$, where $o_j$ is a constant node that takes value $p_j$, $s$ is switches that takes value as: if $q_j=1$, $s$ is pass, if $q_j=0$, $s$ is negation. The circuit with parameters $O_p^q$ has this property: if parameter $p=q$, $O_p^q = 1$, else, i.e. $p \ne q$, $O_p^q = 0$. 

Then we have this circuit: 
$$
V(p) = \bigvee\limits_{q \in \B^J} (C_q \land O_p^q) 
$$
This circuit with parameters $V(p)$ is the boolean circuit with parameters expressing $\varphi(x, p)$. Note, the notation $V(p)$ means: when $p$ is not specified as a fixed parameter, $V(p)$ is circuit with parameters. But, when it is specified, say $p_1$, then $V(p_1)$ is a circuit (since parameter is chosen).

\subsubsection*{Sub-partition Function and Partition Function}
With circuit with parameters $V(p)$, we can have a circuit for a sub-partition function and circuit for partition function. Suppose $P = \{p_1, p_2, \ldots, p_L\}$ is a list of partition vectors, and $SPar_{K,N,P}(x) = \varphi(x, p) \odot P$ is a sub-partition function. Using $V(p)$, we can have a circuit: 
$$
D_P = \bigvee\limits_{p \in P} V(p) 
$$
$D_P$ is the circuit expressing $SPar_{K,N,P}$. If $P$ is the whole partition vector space, then $D_P$ expressing $Par_{K, N}$.  

We can see that $V(p)$ has 2 parts. One is $C_q$. For given $q$, as we discussed above, the size of $C_q$ is proportional to $KN$. Another part is $C_q$ to join with $*O_p^q$. The size of possible $p$ to be chosen is quite big, which is proportional to $2^L$. Questions arise: can we simplify the circuit $V(p)$? Can we reduce the range of choices? Same questions for $D_P$. This is what we are going to discuss in next section.

\section{Complexity of Partition Function}
In this section, we are going to use tools discussed above (FE, PSS, boolean function with parameter and trial-and-error fashion) on sub-partition functions and partition function. We expect that these tools can help us to gain deep insight of the partition function.

Partition function $Par_{N, N}, \ N = 2, 3, \ldots$ is what we are interested. For simplicity, we can use $Par_N$ or just $Par$ for $Par_{N, N}$. So, $Par_N: \B^{N^2} \to \B$. We are also interested in sub-partition functions. $PV_N = PV$ is the partition vector space. For a list of partition vectors $P = \{p_1, p_2, \ldots, p_L\} \subset PV_N$, we have a boolean function: $SPar_{N,P} : \B^{N^2} \to \B$, $SPar_{N,P}(x) = \varphi(x, p) \odot P =  \varphi(x, p) \odot  \{p_1, p_2, \ldots, p_L\}$. If $P$ is the whole partition vector space, then $SPar_{N,P} = Par_N$. 

$PV_N$ is a finite set. Specially, we can make an order of all partition vectors, i.e., all partition vectors can be written in this way: $PV_N =  \{p_1, p_2, \ldots, p_J\}$, where $J = |PV_N| = 2^{N-1} -1$ is the number of all partition vectors. For such order, we can form a sequence of subset of partition vectors: $P_1 = \{p_1\}, P_2 = \{p_1, p_2\}, \ldots$. Generally, $P_j =  \{p_1, p_2, \ldots, p_j\}, j = 1, 2, \ldots, J$. Clearly, $P_1 \subset P_2 \subset \ldots \subset P_J$, and $P_J = PV_N$. For simplicity, we use $Q_j$ to stand for $SPar_{N,P_j}$, so, $Q_j$ is a sequence of boolean functions: $Q_j:  \B^{N^2} \to \B, Q_j(x) = \varphi(x, p) \odot P_j, j = 1, 2, \ldots, J$. We are interested in this sequence of sub-partition functions. Note, $Q_J = Par$.  

We are going to study the sequence of sub-partition function $Q_j$, eventually to reach $Par$. The sequence $Q_j$ of course depends on the choice of the order to partition vector space: $PV_N =  \{p_1, p_2, \ldots, p_J\}$, $J = 2^{N-1} -1$. There are many possible such orders. But, as we will shown below, our results will hold for any such order.  

First, we define some subspace of $\B^{N^2}$ as below:
\begin{equation*}
\begin{split}
& Z \subset \B^{N^2}, Z = \{x \in \B^{N^2} \ | \ Par(x) = 0\}, \\
& W \subset \B^{N^2}, W = \{x \in \B^{N^2} \ | \ Par(x) = 1\}, \\
& W_j \subset \B^{N^2}, W_j = \{x \in \B^{N^2} \ | \ Q_j(x) = 1\}, j=1, 2, \ldots, J
\end{split}
\end{equation*}

Easy to see $Z \cup W = \B^{N^2}, Z \cap W = \emptyset$. We also have following lemma.

\begin{lemma}[\bf Property of $W_j$]
For the sequence of subsets $W_j, j=1, 2, \ldots, J$ defined above, we have: $W_1 \subset W_2 \subset \ldots \subset W_J$, and all these inclusions are true inclusion, i.e. $W_{j+1}\setminus W_j \ne \emptyset, j = 1, 2, \ldots, J-1$, and $W_J = W$. 
\end{lemma}
{\bf Proof:} By definition, for any $j$, $Q_j(x) = \varphi(x, p) \odot P_j$, $Q_{j+1}(x) = \varphi(x, p) \odot P_{j+1}$. Easy to see, if $x \in W_j$, then $Q_j(x) = 1$, thus $Q_{j+1}(x) = 1$, so $x \in W_{j+1}$, that is to say, $W_j \subset W_{j+1}$. From the lemma "Uniqueness of Partition Vector", there is one $x \in \B^{N^2}$, so that $\varphi(x, p_{j+1}) = 1$, and for any other $p$, $\varphi(x, p) = 0$, so $Q_{j+1}(x) =1$ and $Q_i(x) = 0, i=1, 2, \ldots, j$, thus $W_{j+1}\setminus W_j  \ne \emptyset$. Finally, since $Q_J = Par$, $W_J = W$. 
\hfill$\blacksquare$\\

Following, we want to show this fact: for the sub-partition function, $SPar_{N,P} : \B^{N^2} \to \B, SPar_{N,P}(x) = \varphi(x, p) \odot P =  \varphi(x, p) \odot  \{p_1, p_2, \ldots, p_L\}$, a PSS of $SPar_{N,P}$ must have at least one element so that it is unique to each partition vector $p_j, j=1, 2, \ldots, L$.

We start from sub-partition function over only one partition vector.  

\begin{lemma}[\bf One Partition Vector]
Suppose $P =  \{p_1\}$ and $f(x) = SPar_{N,P}(x) = \varphi(x, p) \odot P =  \varphi(x, p) \odot  \{p_1\}$. If $S$ is a PSS of $f$, then there is at least one element $x \in S$ so that $\varphi(x, p_1) = 1$.  
\end{lemma}
{\bf Proof:} Use contradiction. Suppose $S$ has no such element. Thus, for any $x \in S$, $f(x) = \varphi(x, p_1) = 0$, so $f(x) = 0$. Let circuit $C0$ is such a circuit: It is the simplest circuit that for any input, output is always 0. Clearly, $C0$ has zero node, i.e. $d(C0) =0$. So, the circuit $C0$ fits $f$ on $S$, and $d(C0)$ reaches minimum. Since $S$ is PSS of $f$, FE on $S$ to fit $f$ must get a circuit expressing $f$. So $C0$ should express $f$, it means: $\forall x \in \B^{N^2}, f(x) = 0$. But, due to lemma "Uniqueness of Partition Vector", $\exists x \in \B^{N^2}, \varphi(x, p_1) = 1$, so $f(x) = 1$. The contradiction proves lemma.
\hfill$\blacksquare$\\

Note, in above proof, $SPar_{N, P} =  \varphi(x, p) \odot \{p_1\} = \varphi(x, p_1)$. Also, we can note that in the proof, we have not put any restriction on $p_1$. We then consider sub-partition function over only 2 partition vectors. 

\begin{lemma}[\bf Two Partition Vectors]
Suppose $P =  \{p_1, p_2\}$, $f(x) = SPar_{N,P}(x) = \varphi(x, p) \odot P =  \varphi(x, p) \odot  \{p_1, p_2\}$. If $S$ is a PSS of $f$, then there is at least one element $x \in S$ so that $\varphi(x, p_1) = 1, \varphi(x, p_2) = 0$, and also there is at least another element $y \in S$ so that $\varphi(y, p_1) = 0, \varphi(y, p_2) = 1$.   
\end{lemma}
{\bf Proof:} Still use contradiction. Suppose: $S$ has no element so that $\varphi(x, p_1) = 1, \varphi(x, p_2) = 0$. Thus, for any $x \in S$, if $\varphi(x, p_1) = 1$, then must $\varphi(x, p_2) = 1$. Thus, on $S$, if $f(x) = 0$, it must $\varphi(x, p_2) = 0$, if $f(x) = 1$, it must either $\varphi(x, p_1) = 1$ or $\varphi(x, p_2) = 1$, but for either case, must $\varphi(x, p_2) = 1$. This means: on $S$, $f(x) = \varphi(x, p_2)$. Let circuit $C2$ be such a circuit: $C2$ expressing $\varphi(x, p_2)$ and $d(C2)$ reaches minimum. So, the circuit $C2$ fits $f$ on $S$, and $d(C2)$ reaches minimum. Since $S$ is PSS of $f$, FE on $S$ to fit $f$ must get a circuit expressing $f$. Since $C2$ is such a circuit, $C2$ should express $f$, it means: $\forall x \in \B^{N^2}, f(x) = C2(x) = \varphi(x, p_2)$. But, due to lemma "Uniqueness of Partition Vector", there is at least one element $x \in \B^{N^2}$ so that $\varphi(x, p_2) =0$ but $\varphi(x, p_1)=1$, i.e. $f(x) = 1, \varphi(x, p_2) =0$. This is a contradiction, which proves: $\exists x \in S, \varphi(x, p_1) = 1, \varphi(x, p_2) = 0$. 

By the exactly same way, we can prove: $\exists y \in S, \varphi(y, p_1) = 0, \varphi(y, p_2) = 1$.
\hfill$\blacksquare$\\

Note, in above proof, $SPar_{N, P} = \varphi(x, p) \odot \{p_1, p_2\}$. Also, we can note that in the proof, we have not put any restriction on $p_1, p_2$. The only requirement is that $p_1, p_2$ are different partition vectors. Next, we consider more general sub-partition function.

\begin{lemma}[\bf More Partition Vectors]
Suppose $P=\{p_1, p_2, \ldots, p_L\}$, $1 \le L \le J$, and $f(x) = SPar_{N,P}(x) = \varphi(x, p) \odot P =  \varphi(x, p) \odot  \{p_1, p_2, \ldots, p_L\}$. If $S$ is a PSS of $f$, for any given $1 \le i \le L$, then there is at least one element $x \in S$ so that $\varphi(x, p_i) = 1$ and for any other $j$, $1 \le j \le L, i \ne j$, $\varphi(x, p_j) = 0$.   
\end{lemma}
{\bf Proof:} Still use contradiction. Suppose: give a $i$, $1 \le i \le L$, but $S$ has no element so that $\varphi(x, p_i) = 1, \varphi(x, p_j) = 0$ for any other $1 \le j \le L, i \ne j$. It means, for any $x \in S$, if $\varphi(x, p_i) = 1$, then at least there is one $j, 1 \le j \le L, i \ne j$ so that $\varphi(x, p_j) = 1$. 

Let's consider this set of partition vectors: $P' = P\setminus \{p_i\}$, i.e. take $p_i$ out of $P$. And, consider a sub-partition function over $P'$: $g = \varphi(x, p) \odot P'$. 

So, we can see: for any $x \in S$, if $f(x) = 0$, then $g(x) = 0$; for any $x \in S$; if $f(x) = 1$, must $\varphi(x, p_j)=1$, for some $j, 1 \le j \le L$, so there are 2 cases: either  $\varphi(x, p_i)=1$ or  $\varphi(x, p_i)=0$, for former case, $g(x) = 1$ follows, for latter case, if $\varphi(x, p_i)=0$, so must be some $j, 1 \le j \le L, j \ne i$ so that $\varphi(x, p_j) = 1$, it means $g(x)=1$. So, we know: $\forall x \in S, f(x) = g(x)$.   

Let circuit $CI$ be such a circuit: $CI$ expressing $g$ and $d(CI)$ reaches minimum. So, the circuit $CI$ fits $f$ on $S$, and $d(CI)$ reaches minimum. Since $S$ is PSS of $f$, FE on $S$ to fit $f$ must get a circuit expressing $f$. $CI$ is such a circuit, so $CI$ should express $f$, it means: $\forall x \in \B^{N^2}, f(x) = CI(x) = g(x)$. But, due to lemma "Uniqueness of Partition Vector", there is at least one element $x \in \B^{N^2}$ so that $\varphi(x, p_i) =1$ but $\varphi(x, p_j)=0$ for any other partition vector $p_j$, i.e. $f(x) = 1, g(x) =0$. The contradiction proves: $\exists x \in S, \varphi(x, p_i) = 1, \varphi(x, p_j) = 0, \forall i, 1 \le j \le L, i \ne j$.  
\hfill$\blacksquare$\\

Note, in the proof, we have not put any restriction on $p_1, p_2, \ldots, p_L$.

In above lemmas, we have shown: for a sub-partition function, $SPar_{N,P} : \B^{N^2} \to \B, SPar_{N,P}(x) = \varphi(x, p) \odot P =  \varphi(x, p) \odot  \{p_1, p_2, \ldots, p_L\}$, if $S$ is a PSS of $SPar_{N,P}$, for each partition vector $p_j$, $S$ must have at least one element that is special to $p_j$. We put this into a lemma. 

\begin{lemma}[\bf PSS of $Q_j$]
If $S$ is a PSS of $Q_j, 1< j \le J$, $S$ must have at least one element in $W_1$, and at least one element in $W_2\setminus W_1$, \ldots, at least one element in $W_j\setminus W_{j-1}$.  
\end{lemma}
{\bf Proof:} Easy to see: if $\varphi(x, p_1) =1$ and $\varphi(x, p_i) = 0$ for all $i \le j, i \ne 1$, then $x \in W_1$. According to above lemma "More Partition Vectors", $S$ must have at least one such $x$. Also, easy to see: if $\varphi(x, p_2) =1$ and $\varphi(x, p_i) = 0$ for all $i \le j, i \ne 2$, then $x \in W_2\setminus W_1$.  According to lemma "More Partition Vectors", $S$ must have at least one such $x$.  

The other is by the same argument. 
\hfill$\blacksquare$\\. 

The lemma "PSS of $Q_j$" gives one very essential property of $Q_j$. Specially, for $Par = Q_J$, we have: if $S$ is PSS for $Par$, it must have at least one element in $W_1$, and at least one element in $W_2\setminus W_1$, \ldots, $W_j\setminus W_{j-1}$, \ldots, finally, at least one element in $W_J\setminus W_{J-1}$.
 
Note, in the proof, there is no any dependence on any particular order about $PV_N =  \{p_1, p_2, \ldots, p_J\}$. So, the statement we make above: "There are many possible such orders. But, as we will shown below, our results will hold for any such order. " is correct.

We should note: the above lemmas are consequence of lemma "Uniqueness of Partition Vector".  

For "PSS of $Par$, we can see the illustration in Fig. 1.

\begin{center}
\begin{tikzpicture}
    \draw (10cm,0) ellipse (2cm and 0.75cm); 
    \draw (10cm,0) ellipse (2.8cm and 0.8cm); 
    \draw (10cm,0) ellipse (4.0cm and 0.85cm);
    \draw (10cm,0) ellipse (5.5cm and 0.9cm);
    \draw [line width=0.6mm](7.5cm, 0) ellipse (1.8cm and 0.4cm);
	\node at (10cm, 0) {$W_1$};
	\node at (12.4cm, 0) {$W_2$};
 	\node at (13.8cm, 0) {$\ldots\ldots$};
 	\node at (15cm, 0) {$W_J$};
	\node at (7.8cm, 0) {$S$};
\end{tikzpicture} \\
\end{center}
\begin{center}
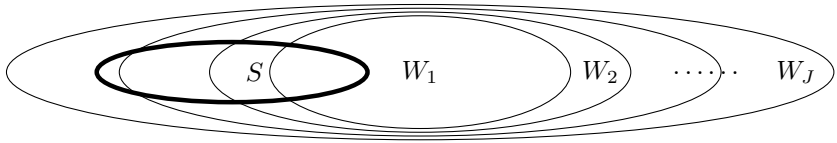
{\bf Fig. 1  Illustration of $S$ and $W_j, j=1, 2, \ldots, J$, $S$ is a PSS of $Par$} 
\end{center}

By this property, we get a lower bound for circuit expressing $Q_j$, and particularly a lower bound for circuit expressing $Par$. 

\begin{lemma}[\bf Size of mPSS]
For sub-partition function $Q_j, j=1,2,\ldots,J$, and suppose $S$ is a mPSS of $Q_j$, then $j < |S|$. Specially, for partition function $Par_N$, if $S$ is a mPSS of $Par_N$, then $2^{N-1} \le |S|$. So, if $C_N$ is a circuit expressing $Par_N$, then $2^{N-1} \le 3d(C_N)$. 
\end{lemma}
{\bf Proof:} By lemma "PSS of $Q_j$", if $S$ is a mPSS of $Q_j$, so it is a PSS, thus, $S$ must have at least one element in $W_1$, at least one element in $W_2\setminus W_1$, \ldots, at least one element in $W_j\setminus W_{j-1}$. That is to say, $S$ has at least $j$ different elements, so $j < |S|$. Particularly, for partition function $Par_N = Q_J$, and $J= 2^{N-1}-1$, so, $2^{N-1} \le |S|$. If $C_N$ is circuit expressing $Par_N$, by the theorem "Circuit implies PSS", there is a PSS $S'$ of $Par_N =Q_j$, and $|S'| \le 3d(C_N)$. Since $S$ is mPSS, we have: $2^{N-1} \le |S| \le |S'| \le 3d(C_N)$.   
\hfill$\blacksquare$\\

This lemma gives lower bound of circuit expressing $Par_N$. We put this into theorem below. 
 
\begin{theorem}[\bf Computational Complexity of Partition Function]
The computational complexity of $Par_{N,N}$ is greater than $\eta 2^N$, where $\eta$ is a constant. 
\end{theorem}
{\bf Proof:} $Par_{N,N} = Par_N$, by the lemma "Size of mPSS", for any circuit $C$ expressing $Par_N$, we have  $2^{N-1} \le 3d(C)$, so $\tfrac{1}{6} 2^N \le d(C)$.  
\hfill$\blacksquare$\\

Now, we understand the complexity of partition function for a special case, i.e. $K=N$, or $L = N^2$: $Par_{N,N}, N = 2, 3, \ldots$, we then turn attention to more general partition function $GPar_L, L = 2, 3, \ldots$. We have following theorem.

\begin{theorem}[\bf Computational Complexity of General Partition Function]
For general partition function $GPar_L, L =2, 3, \ldots$, the computational complexity $C$ is greater than $\eta L 2^{[\sqrt L}]$, where $[w]$ is the integer part of $w$, and $\eta$ is a constant.  
\end{theorem}
{\bf Proof:} Let denote the computational complexity of $GPar_L$ as $\rho_L$. If $L = N^2$. then $GPar_L = Par_{N,N}$, so we have $\eta 2^N < rho_L$, or $\eta 2^{\sqrt L} < rho_L$, where $\eta$ is a constant, and in this case $\sqrt L = [\sqrt L]$. If $N^2 \le L < (N+1)^2$, clearly $\rho_{N^2} < \rho_L$. Then, we have $\eta 2^{[\sqrt L]} = \eta 2^{N} < rho_L$. 
\hfill$\blacksquare$\\

We can discuss another issue. FE on a sampling $S$ to fit $f$ will generate a circuit $C$. The relationship between $f$ and $S$ is very complicated and we can not tell what would be $C$ generally. However, for the partition function, we have following lemma, which gives a surprisingly straightforward answer.

\begin{lemma}[\bf FE to fit Partition Function ]
Suppose $T$ is a PSS of all those boolean functions used in addition and subtraction (as shown in section 4). And, suppose $S \subset B^{N^2}$ is a sampling set and $T \subset S$, then if FE on $S$ to fit partition function $Par_N$ generates a circuit $C$, then $C$ must express $f_t: \B^{N^2} \to \B, f_t(x) = \varphi(x, p) \odot \Psi_S$, where $\Psi_S = \{p_1, p_2, \ldots, p_L\}$ is the partition vector set associated with $S$. 
\end{lemma}
{\bf Proof:} Since $T$ is a PSS of all boolean functions used in addition and subtraction, so is $S$. So, for a given $p$, FE on $S$ to fit the boolean function $z(x) = \varphi(x, p)$, we will get a circuit that expresses $z(x) = \varphi(x, p)$. Thus, FE on $S$ to fit $Par_N$ is equivalent: FE on $S$ to fit $\varphi(x, p) \odot \Psi_S$. This proves lemma.  
\hfill$\blacksquare$


This lemma tells us: if $S$ contains PSS of addition and subtraction (which is very small, relatively), then FE on $S$ to fit $Par_N$ will generate a sub-partition function over the partition vector set associated with $S$. This is a very nice and strong property.

\subsection*{Complexity of Number Partition Problem}
The number partition problem is to answer: whether a set of $N$ natural numbers $\Omega$ can be divided into two subsets $\Omega_1$ and $\Omega_2$ so that the sum of the numbers in $\Omega_1$ equals the sum of the numbers in $\Omega_2$. The size of problem is $N$. In the problem, the number is integer without restriction. In order to put the problem into the framework of boolean functions, we have to put restriction on the size of integer, i,e. $K$-bits integer. So, partition function has 2 sizes: one is the size of problem, $N$, another is the size of integers, $K$. But, the size of problem (i.e. $N$, how many numbers to be partitioned) is same in both the number partition problem and the partition function.  

In previous discussions, we set $K = N$. As discussions in \cite{mertens} indicates that such setting makes partition function interesting. And, $K=N$ makes partition function easier to handle and gives us convenience. In such setting, as theorem 5.6 tells us, the computational complexity of partition function $Par_N = Par_{N,N}$ has lower bound $\eta 2^N$. 

In our discussions, we write the partition function into this form: $Par_N(x) = \varphi_N(x, p) \odot PV_N$, which can help us to see the complexity of the partition function more clearly. The form shows that there are 2 kinds of complexity. One is the complexity to compute $\varphi_N$ once $p$ is given (which are $N-1$ addition/subtraction on $N$-bits integers). Another is the complexity to find the correct partition vector from $PV_N$ whose size is exponential to $N$. It is well known that the first complexity is proportional to $N^2$ (in the examples in last section, we have shown this as well). And, in above discussions (lemma 5.1 to 5.5), we have shown that there is no way to reduce the second complexity to be smaller than $\eta 2^N$. Such form tells us well where the computational complexity comes from. 

The boolean function $Par_N$ is one special case for the number partition problem, since it is just the number partition problem restricting the integer to $N$-bits integer. Any Turing machine (or any algorithm) that can compute the number partition problem, can also compute $Par_N$. Thus, the computational complexity of the number partition problem is higher than the computational complexity of $Par_N$. Since the lower bound for complexity of $Par_N$ is $\eta 2^N$, the computational complexity of the number partition problem with size $N$ also has lower bound $\eta 2^N$. 

Now, we can conclude: the lower bound of computational complexity of the number partition problem is exponential to the size $N$. Here, we quote Cook: "Thus to prove $P \ne NP$ it suffices to prove a super-polynomial lower bound on the size of any family of Boolean circuits solving some specific NP-complete problem, such as 3-SAT." \cite{cook} Thus, we have shown {\bf P} $\ne$ {\bf NP}.

\section{Some Further Thoughts}
In this section, we write down some extended thoughts, which might help us to better understand the tools and methods that we used in this study. 

\subsection{Complexity of Learning vs. Complexity of Computing}
The complexity of learning gauges the efforts to learn a boolean function, which can be measured by the size of mPSS. The complexity of a boolean function is the lower bound of a boolean circuit expressing the boolean function. According to the 2 theorems: PSS implies circuit, and circuit implies PSS, the 2 complexities are equivalent. This is the fundamental thoughts in this study. Such a relationship should be studied further.

\subsection{Contributing Back to Learning Theory}
This research on complexity of partition function gets its inspiration and tools from our studies on universal learning machine. However, we believe that this research can feedback to learning theory and push learning machine to higher level. For example, in current learning theory, the learning target quite often is just a specific boolean function. However, this study suggests that a boolean function with parameters and list of parameters, perhaps, is a much better learning target. To learn a boolean function with parameters could be much more effective and efficient than just a single one. It is the effectiveness of the boolean functions with parameters in this research work suggesting us to look back learning theory and think so. We will continue work in this direction.

\subsection{Canonical Form}
A sequence of boolean functions,  $\{f_N\}, f_N: \B^N \to \B, N = 1, 2, \ldots$, is a powerful computational model, which Avi Wigderson described as "hardware analog of an algorithm" \cite{wigderson}. Sequence of partition functions is a special case of sequence of boolean functions. One key used in current study is: to write the partition function in this form: $f(x) = \varphi(x, p) \odot \Psi$. This form reveals the computational complexity clearly. In this form, there are 2 parts: one is $\varphi$, which is polynomial; another is $\Psi$ that can not be reduced to less, and $\Psi$ is exponential. 

Such a form is the very key of our proof. Naturally, we would ask: Is such a form also available to other sequence of boolean functions?

For a sequence of boolean functions $\{f_N\}, f_N: \B^N \to \B, N = 1, 2, \ldots$, if $f_N$ can be written in this form: $f_N(x) = \varphi_N(x, p) \odot \{p_1, p_2, \ldots, p_T\}$, where $\varphi_N: \B^N \times \B^K \to \B$ is a boolean function with parameters, parameter dimension is $K$, $K=K(N)$ is a function of $N$, the complexity of $\varphi_N$ is polynomial to $N$, and parameter list: $\Psi_N = \{p_1, p_2, \ldots, p_T\}$ is in $\B^K$, and the size of list $T=T(N)$ is a function of $N$, and the list of parameters $\Psi_N$ could not be reduced, i.e. $T$ could not become smaller. We will call such a form $\varphi_N \odot \Psi_N$ as a canonical form for $\{f_N\}$. We put forward a conjecture.

\begin{conjj}[\bf Canonical Form]
For any sequence of boolean functions $\{f_N\}, N=1, 2, \ldots$, it can be written in a canonical form: $f_N = \varphi_N \odot \Psi_N$
\end{conjj}

This is a conjecture. Currently, we do not have further thoughts about proving it or disproving it. It is our current belief that such a form can play some critical roles in complexity. Only further studies can tell us more.

\subsection{Extend to Other Problems}
Could the approach we used in this study be extended to other computational problems, such as SAT, etc? We can summarize the approach as: First, find the canonical form of the problem. Canonical form isolates the complexity out so that it is easier to handle. Such a form is very natural for partition function (almost immediately from its definition). Second, make use of mPSS. In order to do so, we need deep knowledge that is specific to the problem.  For partition function, such knowledge is presented in the lemma "Uniqueness of Partition Vector". 

For a given computational problem, whether or not the above 2 steps can be achieved is questionable. However, as our study suggests, if the 2 steps can be achieved, we might be able to see the insight of the computational problem. Can such an approach form the "new, semantically-interesting ways" \cite{aaronson} that Scott Aaronson talked?

\vskip 0.25in

\subsection*{Acknowledgment}
Special thanks to Dr. Liu, Yu in France. Since 2017, I have had many discussions online with Dr. Liu on Turing machine, Non-deterministic Turing machine and other topics related to computation. These discussions are very insightful and helped me to think in different angle. Thanks to Dr. Huang, Daiyong in Shanghai and Mr. Huang, Chong in Wuhan for various and very useful discussions. Thanks to discussion participants in several WeChat groups, which attract people from whole world together and form a chaotic yet stimulating communication environment for thoughts.


\begin{thebibliography}{99}
\bibitem{paper2} Chuyu Xiong.  Descriptions of Objectives and Processes of Mechanical Learning, arxiv.org, 2017. \\ \htmladdnormallink{http://arxiv.org/pdf/1706.00066.pdf}{http://arxiv.org/pdf/1706.00066.pdf}

\bibitem{paper4} Chuyu Xiong.  Principle, Method, and Engineering Model for Computer Doing Universal Learning (in Chinese), researchage.net, 2018, \\ 
https://www.researchgate.net/profile/Chuyu\_Xiong/research

\bibitem{paper5} Chuyu Xiong. Universal Learning Machine – Principle, Method, and Engineering Model, International Conference of Intelligence Science 2018, Beijing \\ 
https://www.researchgate.net/profile/Chuyu\_Xiong/research

\bibitem{paper8} Chuyu Xiong. Sampling and Learning for Boolean Function, arxiv.org, 2020. \\ \htmladdnormallink{http://arxiv.org/pdf/2001.07317.pdf}{http://arxiv.org/pdf/2001.07317.pdf}

\bibitem{paper9} Chuyu Xiong. A Rudimentary model for Noetic Science (in Chinese), researchage.net, 2019. DOI: 10.13140/RG.2.2.31596.72328

\bibitem{paper10} Chuyu Xiong. Subjectivity of Machine and Its Function (in Chinese), researchage.net, 2020. https://www.researchgate.net/profile/Chuyu\_Xiong/research

\bibitem{karm} Narenda Karmarkar and Richard Karp, The Differencing Method of Set Partitioning, Technical Report UCB/CSD 82/113, University of California at Berkeley: Computer Science Division (EECS), 1980.


\bibitem{mertens} Stephan Mertens, The Easiest Hard Problem: Number Partitioning, arxiv.org, 2003, https://arxiv.org/abs/cond-mat/0310317

\bibitem{hayes} Brian Hayes, The Easiest Hard Problem, American Scientist, Sigma Xi, The Scientific Research Society, vol. 90 no. 2, pp. 113–117

\bibitem{wiki1} Wikipedia. Partition problem, 
\htmladdnormallink{https://en.wikipedia.org/wiki/Partition\_problem}{https://en.wikipedia.org/wiki/Partition\_problem}

\bibitem{wiki2} Wikipedia. NP-completeness, 
\htmladdnormallink{https://en.wikipedia.org/wiki/NP-completeness}{https://en.wikipedia.org/wiki/NP-completeness}

\bibitem{wiki3} Wikipedia. Karp's 21 NP-complete problems, \\
\htmladdnormallink{https://en.wikipedia.org/wiki/Karp\%27s\_21\_NP-complete\_problems}{https://en.wikipedia.org/wiki/Karp\%27s\_21\_NP-complete_problems}

\bibitem{wiki4} Wikipedia. Sample complexity,
\htmladdnormallink{https://en.wikipedia.org/wiki/Sample\_complexity}{https://en.wikipedia.org/wiki/Sample\_complexity}

\bibitem{karp} Richard M. Karp. Reducibility Among Combinatorial Problems, 1972 \\
http://cgi.di.uoa.gr/\~sgk/teaching/grad/handouts/karp.pdf


\bibitem{cook} Stephen Cook, THE P VERSUS NP PROBLEM, 2000. \\ \htmladdnormallink{http://www.claymath.org/millennium/P\_vs\_NP/pvsnp.pdf}{http://www.claymath.org/millennium/P\_vs\_NP/pvsnp.pdf}

\bibitem{crama} Yves Crama and PeterL. Hammer. Boolean Functions Theory, Algorithms, and Applications, Cambridge University Press, 2011

\bibitem{aaronson} Scott Aaronson, P $=?$ NP, 2011. \\ 
https://www.scottaaronson.com/papers/pnp.pdf


\bibitem{wigderson} Avi Wigderson, P, NP and mathematics – a computational complexity perspective,
2006. \\ 
https://www.math.ias.edu/~avi/PUBLICATIONS/MYPAPERS/W06/w06.pdf

\bibitem{kugel} P. Kugel. Thinking may be more than computing, \\
http://citeseerx.ist.psu.edu/viewdoc/download?doi=10.1.1.297.2677\&rep=rep1\&type=pdf

\bibitem{kugel2} P. Kugel. You Don't Need a Hypercomputer to Evaluate an Uncomputable Function, \\
https://www.researchgate.net/publication/
\end{thebibliography}
\end{document}